\documentclass[twocolumn,showpacs,prb,amsmath,amssymb,%
superscriptaddress]{revtex4}

\usepackage{graphicx}
\usepackage{dcolumn}
\usepackage{bm}

\begin{document}

\title{Charge susceptibility in the $t-J$ model}

\author{D.N. Aristov}
\altaffiliation[On leave from ]
{Petersburg Nuclear Physics Institute, Gatchina  188300, Russia.}
\affiliation{
Institut f\"ur Theorie der Kondensierten Materie, Universit\"at Karlsruhe,
76128 Karlsruhe, Germany}
\affiliation{
Max-Planck-Institut f\"ur Festk\"orperforschung, Heisenbergstra\ss e 1,
70569 Stuttgart, Germany}
\author{  G. Khaliullin}
\affiliation{
Max-Planck-Institut f\"ur Festk\"orperforschung, Heisenbergstra\ss e 1,
70569 Stuttgart, Germany}

\date{\today}
\begin{abstract}
Momentum and doping dependence of the static charge susceptibility
$\chi({\bf q})$ in the $t-t'-J$ model is investigated.
Correlations lead to a strongly momentum dependent
renormalization of $\chi({\bf q})$.
The charge susceptibility near $(\pi,\pi)$ region of the Brillouin zone
is strongly suppressed as the hole density $\delta$ is decreased. However,
contrary to naive expectations, $\chi({\bf q})$ around ${\bf q} = (\pi,0)$
and $(0,\pi)$ remains large and practically unchanged at $\delta \sim 0.1-0.5$.
This effect is consistent with a tendency towards low-energy charge
fluctuations with the wave vectors along the $\Gamma-X$ direction,
reported
in earlier studies. Our main finding is that the above trends are
amplified by $J$-driven pairing effects, indicating that the pseudogap
formation may promote the charge inhomogeneity.
The next-nearest hopping $t'$ leads to
weakening of the above momentum-selective renormalizations of $\chi({\bf
q})$.
We analyze the effects of long-range Coulomb interaction,
taking into account a layered structure of cuprates. As an application,
the results are discussed in the context of
bond-stretching phonon softening in hole-doped cuprates. In particular,
a peculiar doping and momentum dependence of the electron-phonon
coupling
constant is found.
\end{abstract}


\maketitle

\section{introduction}

Low energy charge fluctuations and charge ordering becomes a hot topic
in
cuprates. Spatial modulation of the electronic states related to
the local charge and/or bond ordering has been reported
(see Ref.~\onlinecite{Han04} and references therein).
Indirect evidence for the low-energy charge dynamics is obtained
from phonon anomalies induced by hole doping
in cuprates. \cite{Pin05,Fuk05,Rez05}
These experiments motivate a theoretical study
of the charge susceptibility in correlated models.
In general, one expects an overall suppression of the electronic density
fluctuations, hence the related charge susceptibility,
as one approaches the Mott insulating limit by removing
the doped holes. On the other hand, it is also known that correlations
may promote low-energy charge instabilities -- {\it e.g.}, so-called
stripe physics in cuprates and other oxide materials. These seemingly
opposite trends indicate that the renormalization of the charge
susceptibility by strong correlations is quite subtle process.

Previous work on a charge response in $t$-$J$, \cite{Wan91,Geh95,Kha96}
and Hubbard models \cite{Zim97} focused mostly on finite frequency charge
response, $\chi^{\prime\prime}(\omega,{\bf q})$,
and on its frequency-integrated value, {\it i.e.}
a structure factor $N({\bf q})$. These quantities provide an
important information on electron-density fluctuation spectrum.
In particular, Ref.\ \onlinecite{Kha96} presented detailed calculations
of $\chi^{\prime\prime}(\omega,{\bf q})$
within a slave-boson framework. A nontrivial momentum structure
of low-energy excitations has been found. Dressing of the doped-holes by
underlying spin excitations -- a phenomenon well known in the context
of magnetically ordered phase of $t$-$J$ model -- has also been captured
within $1/N$ expansion method for spin-disordered state. The results of
Ref.~\onlinecite{Kha96} are in very good agreement with numerical
data. \cite{Toh95,Ede95}

Surprisingly, a static charge susceptibility
$\chi_{{\bf q}}=\chi'(\omega=0,{\bf q})$ has escaped an attention.
To our knowledge, no detailed discussion of the momentum and doping
dependence of $\chi_{{\bf q}}$ in $t$-$J$ model has thus far been
reported. Meanwhile, this quantity which corresponds to the finite
momentum compressibility contains an important information, {\it e.g.}, about
potential charge instabilities. The aim of this paper is to fill this
gap.

Specifically, we calculate $\chi({\bf q})$ in the $t-t'-J$
model and discuss the doping and spin-pairing effects on $\chi({\bf q})$.
Consistent with known results, we observe that correlations may drive
phase separation at small doping which is however
eliminated by long-range Coulomb forces.
Thus the charge instabilities, if any, are expected at intermediate or
large wave vectors. For large momenta, we find that the correlation
effects are highly anisotropic in a momentum space.
While $\chi({\bf q})$ is suppressed
by a "hole-dilution" effect at certain parts of the Brillouin zone, it
could even be enhanced for ${\bf q}$ along the $\Gamma-X$ direction
[from $q=0$ to $(\pi,0)$ or $(0,\pi)$]. As a result, a featureless
charge susceptibility $\chi^{(0)}({\bf q})\sim$~{\it const}
of noninteracting electrons obtains a strong momentum dependence.
Physically, these observations originate from a nontrivial momentum
structure of low-energy charge excitations found in
Refs.\ \onlinecite{Kha96,Toh95,Ede95}.

The main focus of the paper is to investigate how the above
features in $\chi({\bf q})$ are affected by the $J$-term which
induces a pseudogap in the fermionic dispersion.
Formally, this is done by considering fluctuations in the density
channel taking into account also the pairing fluctuations due to
$J$-interaction. We find that the pairing effects cooperate with
a Gutzwiller constraint and enhance its momentum-selective
renormalization of $\chi({\bf q})$.  Another issue, raised in our
study, is the influence of
the next-to-nearest neighbor hopping $t'$, which is shown
to somewhat weaken the above anomalies in $\chi({\bf q})$.
We also address a question how
the momentum dependence of the compressibility is changed by
the Coulomb interaction. We provide a realistic treatment of the Coulomb
potential accounting both for its long-range character and for the
layered lattice structure of cuprates.

There have been a number of discussions in literature on how the
correlations renormalize electron-phonon coupling.
\cite{Kim91,Kul94,Zey96,Bec96,Koc04,Hua03,Ros04,Cit05}
In Holstein-Hubbard type models (relevant to the problem of oxygen
vibrations coupled to the electron-density) it was found that a peculiar
"forward scattering feature" may develop due to correlations.
We will discuss a connection between this observation and our findings
for the charge susceptibility $\chi({\bf q})$. Related to this issue is
the bond-stretching phonon anomalies in cuprates, which are discussed
in the last part of this paper. This part extends the previous
study \cite{Kha97,Hor05} of the phonon-softening problem by including
the pairing and $t'$ effects.

The rest of the paper is organized as follows. Section II
describes the formalism and discusses $\chi({\bf q})$ in the $t$-only
model. Sections III and IV focus on the effects of $J$ and $t'$ terms,
correspondingly. In Section V, we derive a momentum dependence of
Coulomb potential in the layered lattice structure, and
calculate $\chi({\bf q})$
at presence of these interactions. The last Section VI discusses
renormalization of the electron-phonon coupling by correlations.

\section{the model and formalism}

The $t$-$J$ Hamiltonian is
                \begin{eqnarray}
                 {\cal H}&=& -\sum_{ ij }t_{ij}
                 \tilde c _{i\sigma}^\dagger \tilde c _{j\sigma} - \mu
\sum_i
                 n_i
                 \nonumber \\ &&     +
                 J \sum_{\langle ij \rangle}
                 ({\bf s}_i {\bf s}_j -\frac14 n_i n_j)
                 + \sum_{\langle ij \rangle} V_{ij}
                  n_i n_j .
                 \end{eqnarray}

One of the useful approaches in treating the local constraint on the
fermion occupation number, $\sum _{\sigma} c _{i\sigma}^\dagger c
_{i\sigma}
\leq 1$, is the slave-boson representation $ \tilde c _{i\sigma} =
f_{i\sigma} b_i^\dagger $. The above inequality is then replaced by the
constraint $b_i^\dagger b_i + \sum_\sigma f_{i\sigma}^\dagger
f_{i\sigma} = N/2$, with the physical case  of $N=2$. It is convenient
also to consider the limit of the large number of spin indices
(flavors), $N\gg 1$, since calculations are simplified this way (see,
e.g., Ref.~\onlinecite{Wan91} and  references therein).

Formally, the slave-boson approach implements the constraint of no
double occupancy at the operator level. In the large-$N$ limit, the mean
number of bosons $\langle b_i^\dagger b_i \rangle $ is large and
bosonic amplitude contains a large $c-$number component, a bosonic
condensate. It was noted however, that the phase of each slave
boson is a gauge degree of freedom, which is eliminated from the action
by promoting the local Lagrange multipliers into time-dependent fields.
The remaining degree of freedom for slave-bosons is their real-valued
amplitude, $b_j = r_j e^{i \varphi_j} \to r_j$ and one can formulate
the ``radial-gauge'' representation for slave-bosons.

Below we show how the leading-order results obtained for
the charge susceptibility in the large-$N$ slave-boson approach
\cite{Wan91,Kha96} can be reproduced in a simple manner.
To clarify our approach, we let $J=0$ first, so that only a hopping term
is present. We represent the constrained fermions as

                   \begin{equation}
                   \tilde c _{i\sigma}^\dagger   = c _{i\sigma}^\dagger
                   \sqrt{1-n_i},
                   \end{equation}
with $n_i$ the fermion occupation number, $n_i = \sum_\sigma c
_{i\sigma}^\dagger c _{i\sigma}$.

We are interested in the charge susceptibilty of the system, therefore
we consider small fluctuations of the fermionic density, $n_i$, around
its uniform equilibrium value, $\bar n$. We expand the hopping term up
to the second order in \[\phi_i \equiv \delta n_i/2 = (n_i - \bar n)/2
\;,\] with the factor $2$ introduced for later convenience:

                \begin{eqnarray}
                 {\cal H}_t &=& - \sum_{ ij} t_{ij}
                 c _{i\sigma}^\dagger \left [ (1-\bar n)
                 - (\phi_i + \phi_j)
                 \right. \nonumber \\ && - \left .
                 \frac {(\phi_i - \phi_j)^2}{2(1-\bar n )}
                 \right ]c _{j\sigma}.
                 \end{eqnarray}
In the Fourier  representation the first term above becomes a fermionic
dispersion with the renormalized amplitude,
the second term describes the scattering of the fermions on the
fluctuations of density. The third term has more complicated structure,
for our purposes it is
enough to consider its part, which is diagonal in fermionic momenta,
$\propto c^\dagger_{k1, \sigma} c_{k2 , \sigma} $ with $k_1 = k_2 $.
Then we write

               \begin{eqnarray}
                 {\cal H}_t &\simeq &  \sum_{k\sigma} \xi_k
                 c _{k\sigma}^\dagger c _{k\sigma}
                 \nonumber \\ &&
                 + \sum_{ k,q,\sigma} (t_k+t_{k+q})
                 c _{k\sigma}^\dagger c _{k+q,\sigma} \phi_q
                 \nonumber \\ &&
                 +\sum_{kq\sigma}  \frac {t_{k}-t_{k+q}  }{1-\bar n }
                 c _{k\sigma}^\dagger c _{k\sigma}
                 \phi_q \phi_{-q}.
                 \end{eqnarray}
with
       \begin{eqnarray}
       \xi_k &=&  -(1-\bar n) t_k +\mu , \\
       t_k &=& 2t(\cos k_x +\cos k_y) - 4t' \cos k_x \cos k_y,
       \label{spectrum}
       \end{eqnarray}
henceforth we set $t=1$.
The constraint for $n_i$ to be a number of on-site electrons leads to
the appearance of the local Lagrange multipliers  in the action
                 \[
                 -\mu_q \left(2\phi_{-q} -
                 \sum\nolimits_{k,\sigma}
                  c _{k\sigma}^\dagger c _{k+q,\sigma} \right),
                  \]
so that the scattering term takes the form
               \begin{equation}
               \sum_{ k,q} U_{kq }c _{k\sigma}^\dagger c _{k+q,\sigma},
                 \end{equation}
with  $U_{kq } \equiv \phi_q  (t_k+t_{k+q}) + \mu_q $.
The average number of fermions is defined by
$\bar n = 2 \sum_k n_F(\xi_k)$ with $n_F(x)$ the Fermi factor.
One can integrate out the fermions now, and obtain the effective
low-energy action:

               \begin{eqnarray}
                 {\cal F}_t &\simeq & \sum_q \left  [
                 \phi_q \phi_{-q} (\omega_q - \Pi_2 )
                 \right . \nonumber \\ &&- \left.
                 2\phi_q \mu_{-q}(\Pi_1+1) - \mu_q\mu_{-q} \Pi_0\right
],
                 \label{freeEn1}
                 \\
                 \omega_q &=&
                 2 \sum_{k} \frac{t_k- t_{k+q}}{1-\bar n }  n_F(\xi_k)
                 \label{def-omega0},
                 \\
                 \Pi_n&=& \sum_k \frac{(t_k
+t_{k+q})^{n}}{\xi_{k+q}-\xi_k}
                 (n_ F(\xi_k)- n_F(\xi_{k+q})).
                 \label{defPi0}
                 \end{eqnarray}
Requiring the zero variation,  $\delta {\cal F}_t/\delta \mu_q = 0$, we
determine the values of Lagrange multipliers
                 \begin{equation}
                  \mu_q=  - \phi_q (1+\Pi_1)/\Pi_0,
                   \end{equation}
Inserting these values into (\ref{freeEn1}), and recalling that $\phi_q
= n_{q}/2$ at $q\neq 0$, we find
               \begin{eqnarray}
                 {\cal F}_t &\simeq &  \sum_q
                 \frac{n_q  n_{-q}}{4} \left (\omega_q - \Pi_2
                  +\frac {(1+\Pi_1)^2}{\Pi_0}
                 \right).
                 \label{freeEn2}
                 \end{eqnarray}
The above equation for the free energy in the harmonic approximation
should be
compared to the general expression ${\cal F} =  \sum_q (2\chi_q)^{-1}n_q
n_{-q} + \ldots $ with the static charge susceptibility $\chi_q$. The
value of $\chi_q$ as determined  from (\ref{freeEn2}) coincides with the
result of Ref.~\onlinecite{Kha96} in the limit of $\omega=0$ [however, a
so-called polaron correction due to the higher $1/N$-term \cite{Kha96}
is absent in Eq.(\ref{freeEn2})].

Let us first qualitatively analyze the above formula for the charge
susceptibility. We introduce the doping level, or the concentration of
holes,
\[\delta \equiv 1-\bar n, \]
and extract this factor from
the dispersion, $\xi_{{\bf q}}$, and the chemical potential, $\mu$,
 \begin{equation}
\xi_{{\bf q}}=(- t_k + \tilde \mu)\delta
= \xi^{(0)}_{{\bf q}} \delta \;.
 \end{equation}
Note that $\tilde \mu$ is positive and proportional
to the doping level at $t'=0$.

After some straightforward rearrangements of Eq.(\ref{freeEn2}),
the expression for $\chi({\bf q})$ can be represented in the following
form:
\begin{eqnarray}
\chi_{{\bf q}}=\chi^{(0)}_{{\bf q}}
\frac{\delta}{(\delta-\alpha_{{\bf q}})^2 +
2(\tilde\mu \delta + \beta_{{\bf q}})\chi^{(0)}_{{\bf q}}} \;.
\label{Gutzwiller}
\end{eqnarray}
Bare quantities $\chi^{(0)}_{{\bf q}}$, $\alpha_{{\bf q}}$ and
$\beta_{{\bf q}}$ are given in terms of the "noninteracting" dispersion,
$\xi^{(0)}_k$ as follows :
\begin{equation}
\chi^{(0)}_{{\bf q}}=
2 \sum_k \frac{1}{\xi^{(0)}_{k+q}-\xi^{(0)}_k}
                 (n_ F(\xi^{(0)}_k)- n_F(\xi^{(0)}_{k+q})),
 \end{equation}
corresponding to the bare susceptibility, and
\begin{eqnarray}
\alpha_{{\bf q}} &=&
\sum_k \frac{\xi^{(0)}_{k+q}+\xi^{(0)}_k}{\xi^{(0)}_{k+q}-\xi^{(0)}_k}
                 (n_ F(\xi^{(0)}_k)- n_F(\xi^{(0)}_{k+q})),
\label{alpha}
\\
\beta_{{\bf q}}& =&
-\sum_k \frac{\xi^{(0)}_{k+q} \xi^{(0)}_k}{\xi^{(0)}_{k+q}-\xi^{(0)}_k}
                 (n_ F(\xi^{(0)}_k)- n_F(\xi^{(0)}_{k+q})).
\label{beta}
\end{eqnarray}
One can show that the function $\beta_{{\bf q}}$ in (\ref{beta}) is
positive.
Eq.(\ref{Gutzwiller}) should be understood as a renormalization of the
charge susceptibility by correlation effects,
$\chi_{{\bf q}}/\chi^{(0)}_{{\bf q}}=G_{{\bf q}}(\delta)$, where a
function
$G_{{\bf q}}(\delta)$ is given by a fraction in Eq.(\ref{Gutzwiller}).
This momentum and doping dependent factor results from the action
of the Gutzwiller constraint in the density channel.
Whereas a noninteracting susceptibility $\chi^{(0)}_{{\bf q}}$ is
a featureless function in the absence of nesting (at $\delta=0$,
$t'=0$),
the correlation effects bring a pronounced momentum structure
in $\chi_{{\bf q}}$ via the function $G_{{\bf q}}(\delta)$.

At large doping levels, $G_{{\bf q}}(\delta)$ eventually approaches
unity.
The action of this factor at small doping is highly momentum-selective.
Inspecting Eqs.(\ref{Gutzwiller}-\ref{beta}) at $t'=0$, one finds that
$G_{{\bf q}}(\delta)\propto \delta$ for ${\bf q} \sim (\pi,\pi)$
(with omitted logarithmic corrections).  This is simply understood
as a reduction of density fluctuations due to a removal
of the holes. (Alternatively, one may say, that the checkerboard
structure in positions of small amount of doped holes is the
least energetically favorable.) At small momenta, however,
the effect is opposite and one has  $G_{{\bf q}}(\delta)\propto
1/\delta$.
This means a divergent compressibility as one approaches
the Mott limit, in accordance with previous studies of Hubbard
\cite{Fur92}
and $t$-$J$ models, \cite{Koh97,Tan99,McK01} and reflects a well-known
tendency towards phase separation. \cite{Hel97,Dee94,Bec00} Competition
between these two effects -- a hole dilution and phase
separation -- leads to a nontrivial momentum structure.
It is interesting to note that this structure is
complementary to that in the spin sector, where correlations enhance
the spin susceptibility at ${\bf q} \sim (\pi,\pi)$, \cite{Dee94,Win98}
but not at small momenta.

\begin{figure}[tp]
\includegraphics[width=8cm]{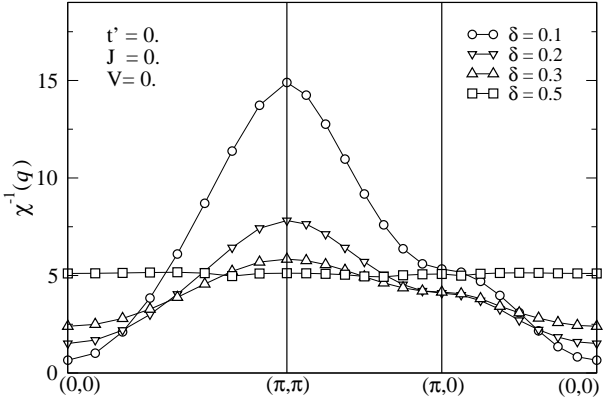}%
\caption{The inverse non-uniform compressibility calculated for
tight-binding spectrum.
\vspace*{0.6cm}
\label{fig:1}}
\end{figure}

\begin{figure}[tp]
\includegraphics[width=8cm]{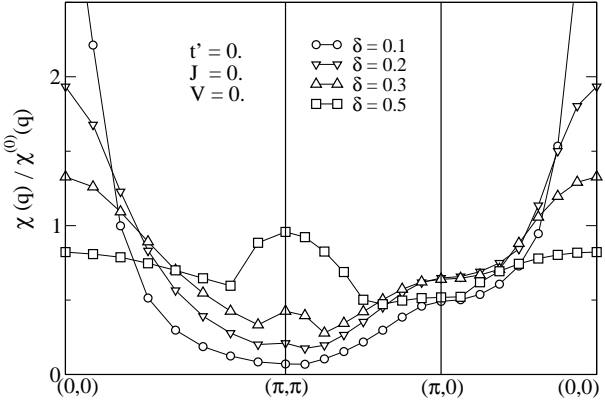}
\caption{Renormalization of the susceptibility due to the
Gutzwiller constraint.
\label{fig:2}}
\end{figure}

\begin{figure}[tp]
\includegraphics[width=8cm]{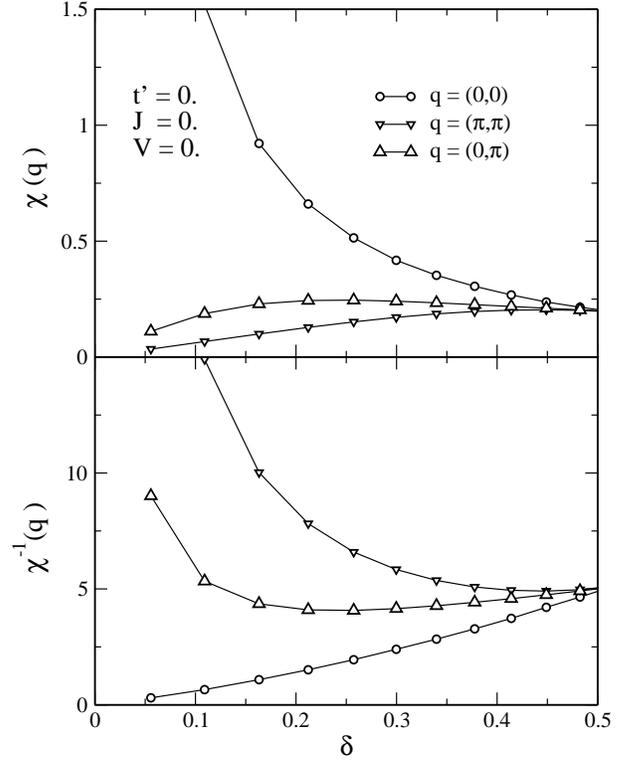}
\caption{Doping dependence of the charge susceptibility (upper panel)
and its
inverse value (lower panel) at the symmetry points.
\label{fig:3}}
\end{figure}

These qualitative observations are further illustrated by the numerical
calculations. In Fig.\ \ref{fig:1}, we show the momentum dependence of
$\chi_{{\bf q}}$ along the symmetry lines in the Brillouin zone.
The above behavior of $\chi_{{\bf q}}$
with doping is clearly visible at the symmetry points.
In order to emphasize this nontrivial momentum dependence induced by
the Gutzwiller constraint, we plot
$\chi_{{\bf q}}/\chi^{(0)}_{{\bf q}}=G_{{\bf q}}(\delta)$
in Fig.\ \ref{fig:2} for several dopings $\delta \simeq (0.1, 0.2, 0.3,
0.5)$.
To see the doping dependence in more detail, we show $\chi_{{\bf q}}$ in
Fig.\ \ref{fig:3} as a function of doping at three symmetry points
${\bf q}=0$, $(\pi,0)$ and $(\pi,\pi)$.
One observes that the curves for $\chi_{{\bf q}}\neq 0$ eventually
turn down at small doping $\delta$.
Remarkably, the value of $\chi_{{\bf q}}$ at $(\pi,0)$ upon decreasing
$\delta$ is somewhat enhanced before the downward turn, which shows the
competition between two trends: phase separation and hole-dilution.

Small momentum anomalies are eliminated in reality by Coulomb
repulsion. A detailed study of this problem is presented in Section V.
We show there that a pronounced momentum structure of $\chi({\bf q})$
with very different doping dependence at $(\pi,\pi)$ and  $(\pi,0)$
regions still remains at the presence of Coulomb interactions.

\section{$J$-term: pseudogap effects}

We consider now how the above observations change at the presence of
pseudogap effects induced by $J$-term in the Hamiltonian. In the spirit
of large-$N$ slave-boson theories, we refer to pseudogap as a fermionic
gap arising from mean-field decoupling of the superexchange interaction
in the pairing channel. A question addressed here is that how such a
gap and fluctuations of the pairing field around uniform mean-field
solution will affect the charge susceptibility.

The four-fermion $J$-term can be represented in the form

                   \begin{equation}
                   {\cal H}_J =
                   -\frac12 \sum_{k1,k2,k3}
                   c_{k1,\uparrow}^\dagger c_{k2,\downarrow}^\dagger
                   c_{k3,\downarrow} c_{k4,\uparrow} (J_{k4-k1} +
J_{k3-k1}),
                   \end{equation}
with $k_4 = k_1+k_2-k_3$ and the nearest-neighbor interaction $J_k = 2J
(\cos
k_x + \cos k_y) $.  Introducing the quantity

        \begin{equation}
        \eta_q^\pm = \sum_{k}
        c_{k+q/2,\downarrow}c_{-k+q/2,\uparrow}\gamma_k^\pm        ,
         \end{equation}
with
          \begin{equation}
           \gamma_k^\pm = (\cos k_x \pm \cos k_y)/2,
           \end{equation}
we represent the $J-$term as
         \begin{equation}
         {\cal H}_J =
         -4J \sum_{q,\alpha =\pm} (\eta_q^\alpha)^\dagger \eta_q^\alpha
        \end{equation}

This expression can be decoupled by the Hubbard-Stratonovich
transformation as
follows
                  \begin{equation}
                   {\cal H}_J =\sum_{q,\pm}
                   \left [ (  d_{q}^\pm \eta_q^\pm   +h.c.)
                   + \frac {| d_{q}^\pm|^2 }{4J}
                   \right ].
                   \label{SC1}
                   \end{equation}
Here $d_{q}^-$ and $d_{q}^+$ stand for the amplitudes
of the $d-$wave and extended $s-$wave pairing, respectively,
in the channel with non-zero total momentum, $q$.

The total Hamiltonian is then quadratic in fermions, which interact with
the fluctuations $\phi_q, \mu_q $  and $d_{q}^\pm$.

We assume that the d-wave pairing sets in, which corresponds to the
non-zero value of the order parameter $d_{0}^- \equiv \Delta$.
The spectrum is given by
$\varepsilon_k^2 =\xi_k^2 + \Delta_k^2$, where
$\Delta_k = d_{0}^- \gamma_k^- $.
The self-consistency gap equation in the small-coupling limit reads
           \begin{equation}
            1= 2 J \sum\limits_k \frac{(\gamma_k^-)^2} {\varepsilon_k
            }\tanh\frac{\varepsilon_k}{2T}.
            \label{gap}
            \end{equation}

Restricting our consideration by the quadratic terms in the above
non-uniform
fluctuations, we would like to obtain an expression similar to
(\ref{freeEn2}), but in the presence of the pairing.
After some standard analysis, we arrive at the bosonic-type action of
the form

          \begin{eqnarray}
          {\cal F}_{t-J} &=& \sum_q \Phi^\dagger_q M_q \Phi_q
          \label{freeEn3a}\\
          \Phi^\dagger_q &=&
          (\phi_q, \mu_q , d_q^+, d_q^- ),
          \label{freeEn3}
          \end{eqnarray}
with the matrix
          \begin{equation}
          M_q\!=\!
          \begin{pmatrix}
          \omega_q\!-\!\Pi_2 &-1\!-\!\Pi_1 & A_1 ^+ &A_1^- \\
          -1\!-\!\Pi_1 & -\Pi_0   & A_0 ^+ &A_0^- \\
          A_1 ^+ &A_0^+ & (4J)^{-1}\!-\!D_{ss} & -D_{sd}  \\
          A_1 ^- &A_0^- & -D_{sd}&  (4J)^{-1}\!-\!D_{dd}
           \end{pmatrix}\!.
           \label{defM}
          \end{equation}

Here in the low-temperature limit
          \begin{eqnarray}
          \omega_q &=&
          \sum_{k} \frac{t_k- t_{k+q}}{1-\bar n }
          \frac{\varepsilon_k-\xi_k}{\varepsilon_k},
          \nonumber
          \\
          \Pi_n&=& \sum_k (t_+ +t_-)^{n}
          \frac{\varepsilon_+ \varepsilon_- -\xi_+\xi_-
          +\Delta_+ \Delta_-}
          {2 \varepsilon_+\varepsilon_-( \varepsilon_+ +
\varepsilon_-)},
         \label{defPi}
          \end{eqnarray}

          \begin{equation}
          \begin{pmatrix}
           A_n^+ \\ A_n^-
           \end{pmatrix}
           = \sum_k (t_+ + t_-)^{n}
           \frac{\Delta_+\xi_- + \xi_+ \Delta_-}
           {2 \varepsilon_+\varepsilon_-( \varepsilon_+ +
\varepsilon_-)}
           \begin{pmatrix}
           \gamma_k^+ \\  \gamma_k^-
           \end{pmatrix},
           \label{defA}
          \end{equation}
and
         \begin{equation}
             \begin{pmatrix}
          D_{ss} \\            D_{sd}  \\            D_{dd}
           \end{pmatrix} =
           \sum_k  \frac{\varepsilon_+ \varepsilon_- +\xi_+\xi_-}
           {2 \varepsilon_+\varepsilon_-( \varepsilon_+ +
\varepsilon_-)}
          \begin{pmatrix}
           [\gamma_k^+]^2 \\    \gamma_k^+ \gamma_k^-     \\
[\gamma_k^-]^2
           \end{pmatrix},
           \label{defD}
          \end{equation}
and we used the shorthand notation $\xi_\pm = \xi_{k\pm q/2}$,
$\varepsilon_\pm = \varepsilon_{k\pm q/2}$,  etc.

We are interested in the charge susceptibility, which should
basically be determined by integrating out the pairing fluctuations in
(\ref{freeEn3}) and setting the Lagrange multipliers $\mu_q$ to their
saddle-point values. Both steps are essentially the same for the
quadratic action, so that the needed compressibility is given by
the upper left element of the inverse matrix $M_q$, namely
 \begin{equation}
 \chi_q = 2  [M_q^{-1}]_{11}.
 \label{invM}
 \end{equation}


Let us briefly discuss here how the inter-site Coulomb interaction
$V_{ij}$ is included in our formalism. We write
it in the form $\sum _{\langle ij \rangle } V_{ij}  (n_i-\bar n)(n_j -
\bar n )
= \frac 12 \sum_q V_q  n_q n_{-q} = 2 \sum_q V_q  \phi_q \phi_{-q}$,
and see that this interaction modifies the
only matrix element in Eq. (\ref{defM}), so that
               \begin{equation}
               [M_q]_{11}  \to \omega_q-\Pi_2 + 2 V_q.
               \end{equation}
The last equation shows that we treat the interaction
$V_q$ within the RPA scheme. It can also be shown that
              \begin{equation}
                \chi_{V,q}^{-1} =
                 \chi_{V=0,q}^{-1}+  V_q,
              \label{chiRPA}
               \end{equation}
i.e.\ one can calculate $\chi_q$ at $V_q=0$, and
include $V_q \neq 0$ afterwards. At the same time, the inclusion
of $V_q \neq 0$ should be done from the beginning for the calculation
of $d-$wave susceptibility, $\chi_{dd}({\bf q})$, see below.

\begin{figure}[tp]
\includegraphics[width=8cm]{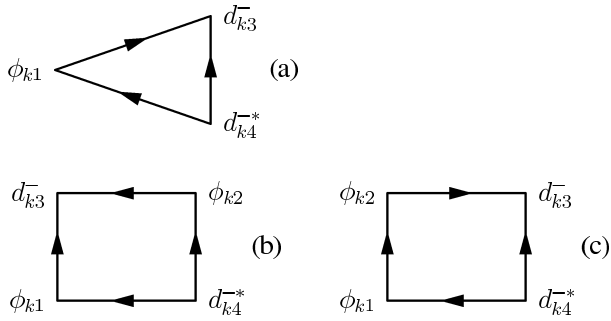}
\caption{
Multi-tail fermionic Feynman diagrams, leading to higher-order
terms in the bosonic action. Fermionic Green functions shown
by lines with arrows, bosonic fields $\phi, d^-$ stand in the vertices.
For simplicity we do not show the internal momenta.
\label{fig:diag}}
\end{figure}

It is worth also noting that in the absence of pairing,
$\Delta_k\equiv 0$, the coefficients
$D_{\alpha\beta}$ in (\ref{defM}) remain finite, and  $A_n^\pm$ vanish.
It means that in the harmonic approximation the superconducting-type
fluctuations affect the charge susceptibility only at finite $\Delta$.
At the same time, in more general treatment, the superconducting
fluctuations $d^\pm$ affect the density fluctuations in the higher
orders  even at $\Delta_k\equiv 0$. Considering multi-tail fermionic
loops, one obtains, e.g., the terms in the bosonic action of the
form~: \[ A^{--}_1 \phi_{k1} d^-_{k2}d^{-\ast}_{k3} +
B^{--}\phi_{k1}\phi_{k2} d^-_{k3}d^{-\ast}_{k4} , \]
etc., where the coefficients $A^{--}_1,  B^{--}$ are defined by the
Feynman diagrams shown schematically in Fig.\ \ref{fig:diag}(a) and
Fig.\ \ref{fig:diag}(b,c), respectively. One can notice that the
diagram in Fig.\ \ref{fig:diag}(b) corresponds to an analog
of Maki-Thompson contribution to paraconductivity and the
diagram in Fig.\ \ref{fig:diag}(c) reflects the "density of states
correction", see Ref.\ \onlinecite{Lar01}. The contribution of the
diagram
Fig.\ \ref{fig:diag}(a) is zero at $\Delta_k\equiv 0$. However,
in the presence of the pairing condensate, one external bosonic
field, $d^-$, sets to a constant, contributing in the lowest
order to the term $A_1^-$ in (\ref{defM}), etc. The calculation
of such fluctuation corrections to the charge susceptibility
is clearly beyond the scope of the present study.

In our numerical calculations we set $J=0.3$, and determined $\Delta$
from the self-consistency equation (\ref{gap}) at $T=0$.
One should note, that at $T=0$ Eq.\ (\ref{gap}) has a
solution $\Delta\neq 0$ for any doping,
although $\Delta$ can be exponentially small. In our calculation,
we ruled out the solutions with  $|\Delta|< 10^{-3}t$,
thus implicitly setting the temperature to be very small but finite,
$T \sim 10^{-3}t$. The matrix $M$ is then found from (\ref{defPi}),
(\ref{defA}), (\ref{defD}), and $\chi_q $ from (\ref{invM}).

\begin{figure}[tp]
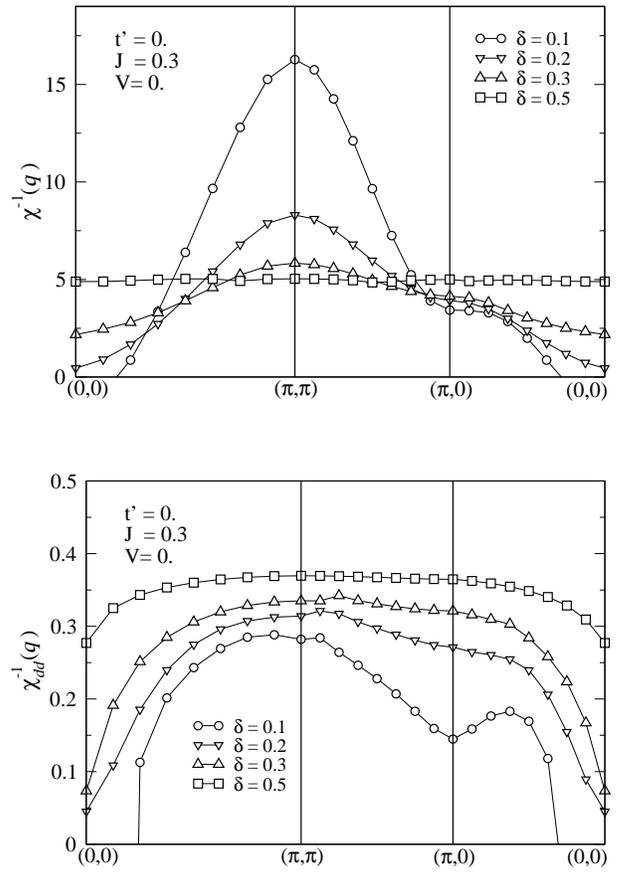

\includegraphics[width=8cm]{fig4a.eps}\vspace*{0.9cm}
\includegraphics[width=8cm]{fig4b.eps}
\caption{
Behavior of inverse charge (upper panel) and d-wave pair (lower panel)
susceptibilities at $J=0.3$.
\vspace*{0.4cm}
\label{fig:4}}
\end{figure}

The obtained results are shown in Fig.\ \ref{fig:4}. Comparing the
upper panel in Fig.\ \ref{fig:4} with the previous Fig.\ \ref{fig:1},
we observe that $\chi_q$ is still rather flat in the
whole Brillouin zone at large dopings. At smaller dopings the above
momentum-selective features are enhanced by the pseudogap appearance,
with somewhat increase of $\chi_q^{-1}$ at ${\bf q}=(\pi,\pi)$ and a new
qualitative change at small wave-vectors. Namely, the compressibility
attains negative values at finite  $\delta\sim 0.15$, which means
that the Gaussian action $\sim  \chi_q^{-1}(\delta n_q)^2$ is unstable
in rather extended range of dopings and the analysis of the next
orders in $\delta n_q$ is needed. This instability is accompanied by the
divergence in the d-wave susceptibility, $\chi_{dd}$,
defined as the $\frac12 [M^{-1}_q]_{44}$, as is seen in the lower panel
in Fig.\ \ref{fig:4}. We remind that the latter quantity is always
positive $\chi_{dd}^{-1} \sim \Delta^2/t$ in the absence of feedback
from density fluctuations to the superconducting ones,
i.e. when $A_n^\pm \equiv 0$.  The unstable pairing part of the action
$\sim \sum_q \chi_{dd}^{-1}({\bf q}) |d_q^-|^2$ particularly means
that the uniformly paired ground state determined by the gap
equation (\ref{gap}) is no longer justified.

Our finding that J-pairing fluctuations and charge fluctuations grow up
concomitantly could be understood as a dymanical modulations of pairing
amplitude consistent with the results by Vojta {\em et al.}
\cite{Voj99,Voj00} It is also noticed that a dramatic enhancement
of the charge susceptibility at small momenta due to $J$ term is
consistent
with previous reports (see, {\it e.g.} Fig. 3 of Ref.\onlinecite{Dee94})
that
the superexchange interaction increases a tendency towards
phase separation. Close to such instabilities, the higher order
terms (beyond the Gaussian action) should be included in the
theory, which problem deserves a separate study.

\section{next-nearest hopping}

We discuss now the effects of next-to-nearest neighbor hopping $t'$
which is present in cuprates and has in fact been suggested
to be a key empirical parameter for superconductivity. \cite{Pav01}
Effects of $t'$ on physical quantities such as spin and fermionic
excitations in the $t$-$J$ model has been found to be substantial, see,
{\it e.g.} Ref.\ \onlinecite{Toh04}. Concerning the charge
compressibility, several studies found that $t'$ hopping
reduces a tendency towards phase separation. \cite{Dee94,Kot04,Whi99}

Consider first the qualitative effect of $t'\neq 0$ in the absence of
$J$ term. Using (\ref{Gutzwiller}), one can still show that the
renormalization factor $G_{\bf q}(\delta) \sim \delta$ at
${\bf q}  = (\pi, \pi)$. At the same time,
for $t' \neq 0$ the chemical potential $\tilde \mu$ does not vanish when
$\delta\to 0$ and we have finite $G_{\bf q}(\delta) \sim 1/\tilde \mu$
at small momenta. It means that finite ({\it positive}) $t'$ reduces the
tendency to the phase separation at small doping, consistent with
previous work. It is interesting to note that the {\it negative} $t'$
has an opposite effect. Indeed, in this case the chemical potential
$\tilde \mu$ is also negative. By inspecting Eq.(\ref{Gutzwiller})
one observes that this may lead to a negative values of the
susceptibility at small momenta and doping, indicating an instability
of the uniform state for $t'<0$ case at small dopings.

\begin{figure}[tp]
\includegraphics[width=8cm]{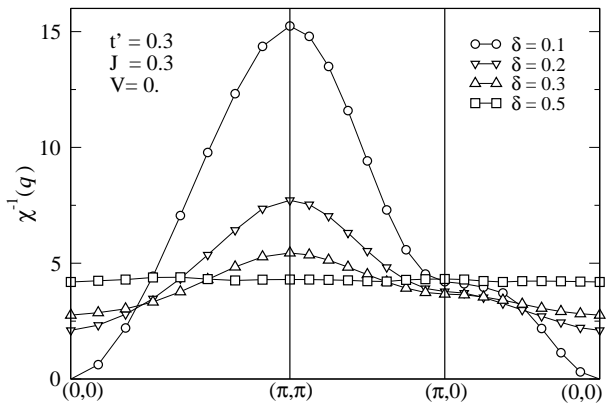}
\caption{
Inverse charge susceptibility at $t'=0.3$ and  $J=0.3$. Anomalies at
small
momenta seen in Fig. 5 are suppressed.
\label{fig:5}}
\end{figure}

We showed in the previous section that the inclusion of $J$ term drives
the system
closer to the instability point for charge fluctuations at $t'=0$,
and the same thing should happen when
next-to-nearest neighbor hopping is present. To verify it,
we recalculated $\chi_{{\bf q}}$ for $t'=0.3$, $J=0.3$ for the
same values of doping $\delta$ as above. The results are shown
in Fig.\ \ref{fig:5}. Comparing it to Fig.\ \ref{fig:4}, one confirms
that the finite next-nearest hopping $t'>0$ somewhat stabilizes the
charge
fluctuations. The comparison to Fig.\ \ref{fig:1} shows, however,
that the effect of the $J$ term is still dominant and $\chi_q$ is
(nearly) divergent at $\delta =0.1$.

We should emphasize that the above statements on the role of $t'$ and
$J$ terms do not explicitly rely on the one-particle properties of the
spectrum, such as van Hove singularities and flat parts of dispersion
around $(0,\pi)$ points. Our discussion includes two-particle Green's
functions, both particle-hole and particle-particle fermionic loops,
and the eventual integration over bosons in the effective action, i.e.\
obtaining Eq.\ (\ref{invM}) from Eq.\ (\ref{freeEn3a}), corresponds
to simultaneous resumming of RPA series in both Gutzwiller and $J$-term
channels.

Summarizing here, the static
susceptibility shows no structure in the Brillouin zone
at large value of the hole doping, $\delta \simeq 0.5$. This
could be expected for a system without strong correlations and
with a large Fermi surface, since in the 2D Fermi gas $\chi_q =
const $ at $q<2k_F$. The flat shape of $\chi_q$ at large $\delta$ is
rather insensitive to the values of the second hopping and
pairing. At smaller dopings, $\chi_q$ demonstrates a pronounced
structure in $q$-space; the tendency to long-scale phase separation is
somewhat weakened by finite values of the second hopping,
but the pairing fluctuations, induced by the $J$ term,
dominate and eventually make the system unstable both in charge
and pairing channels.

\section{long-range Coulomb interaction}

Charge susceptibility is strongly influenced by a non-local repulsion
between the holes. Quite often (in numerical studies in particular)
these interactions are approximated by a nearest-neighbor potential
$V_1$,
which is already sufficient to observe the suppression of phase
separation
effects discussed above. We are however interested in a more detailed
momentum dependence of $\chi_q$. For this purpose, one has to use more
realistic, {\it i.e.} long-range form of the Coulomb potential. We
consider first its momentum dependence in a layered cuprate structure,
taking into account a discrete nature of the lattice within the planes
as well.

At small momenta, where no lattice structure is relevant, a continuum
limit applies:
\begin{equation}
V_C({\bf Q}) = \frac{4\pi e^2}
{\epsilon_{ab}q^2 + \epsilon_c q_z^2} \;.
\label{Coulomb-media-Q}
\end{equation}
Here, $\epsilon_{ab}$ and $\epsilon_c$ are zero-frequency dielectric
constants determined usually from optical data, and $q^2=q_x^2+q_y^2$.
($a$ and $b$ directions are assumed to have the same $\epsilon$).
We use a notation ${\bf Q} = ({\bf q}, q_z)$, with in-plane and
out-of-plane
components, ${\bf q}$ and $q_z$, respectively. For $q_z=0$, the above
equation gives $V({\bf q})=4\pi e^2/\epsilon_{ab}q^2$, a conventional
3D potential. We recall that in 2D (the case of infinitely separated
planes)
$V({\bf q})\propto 1/q$, and we discuss the crossover between
these two regimes below.

We argue here that for our analysis it is possible to neglect
a momentum dependence of the dielectric constants, because at low energy
they are mostly contributed by (dispersionless) optical phonons and
nearly localized, high-energy electronic processes. In this case the
real
space representation of (\ref{Coulomb-media-Q}), valid up to
interatomic distances, reads as follows:
\begin{equation}
V_C({\bf R}) = \frac{e^2}{\sqrt{\epsilon_{ab}\epsilon_c}}\;\;
\frac{1}{[(\epsilon_{ab}/\epsilon_c)z^2 + r^2]^{1/2}}\;.
\label{Coulomb-media}
\end{equation}
Here, $r$ is a distance within $ab$-plane, and $R^2=r^2+z^2$. In
isotropic
case, $\epsilon_{ab}=\epsilon_c$, a familiar expression $e^2/\epsilon R$
follows from this equation.


Let us consider now a lattice with periodicity $a$ within the planes and
$d$ along the $c$-axis (in La$_2$CuO$_4$-structure $d$ is a half of the
$c$ axis lattice parameter, i.e. $d=c/2$). We determine the Coulomb
repulsion $V_{ij}$ between the electrons, referring to the sites $i$ and
$j$.
In our tight-binding situation the electronic wave-functions are
almost localized around the $i$-th ion and their amplitude squared gives
the density around this ion. We denote this density, or charge
distribution
function, by $f({\bf R} - {\bf R}_i)$, and write
    \begin{eqnarray}
    V_{ij}&=&\int d{\bf R}' d{\bf R}'' f({\bf R}' - {\bf R}_i) f({\bf
R}''
     - {\bf R}_j)
    V_C({\bf R}' -{\bf R}'' )
     \label{Vij} \nonumber \\
     &=&\int \frac{d{\bf Q}}{(2\pi)^3} |f({\bf Q})|^2 V_C({\bf Q})
     e^{i {\bf Q} ({\bf R}_i - {\bf R}_j) } \;,
    \end{eqnarray}
where the integration is over the whole continuum and  $V_C$ is
given by Eqs. (\ref{Coulomb-media-Q}) and (\ref{Coulomb-media})
in a momentum and real spaces, respectively.
The Fourier transform of $V_{ij}$ then reads as
    \begin{eqnarray}
    V({\bf Q})&=&\sum_{j\neq i} V_{ij} e^{i {\bf Q} ({\bf R}_i - {\bf
R}_j)}
    \\
    &=&\frac{1}{a^2 d} \sum_{{\bf G}_{3}} |f({\bf Q} + {\bf G}_{3})|^2
    V_C({\bf Q} + {\bf G}_{3} ) - V_{ii}
\nonumber
    \label{gen-Cou}
    \end{eqnarray}
with ${\bf G}_{3}$ the 3D wave vector of reciprocal lattice.

On the physical grounds, one expects that a doped hole
(the Zhang-Rice singlet) is a rather extended object in the $ab$-plane
and nearly localized in this plane. A reasonable choice for a hole-shape
function is thus
\[f({\bf R})=(\kappa^2/2\pi) e^{-\kappa r} \delta(z) \] which decays
at distances $1/\kappa$ in the plane. Physically, the size of the
Zhang-Rice singlet should at least be about Cu-O distance,
so $\kappa \sim 2/a$ might be a representative value.
A momentum counterpart of the latter function,
\begin{equation}
f({\bf Q})= (1+q^2/\kappa^2)^{-3/2}
\label{fq}
\end{equation}
should then be understood as a formfactor of the Zhang-Rice singlet.

Given that the formfactor $f({\bf Q})$ is independent of the $q_z$
component,
the summation over $G_z = 2\pi n/d$ ($n=0,\pm1,\pm2, \ldots$) in
(\ref{gen-Cou}) is easily performed for any $q_z$.
The result for our primary case of interest, $q_z=0$, is
    \begin{eqnarray}
    V_ {\bf q} \! \equiv \! V({\bf q},0) \!= \sum_{{\bf G}_2}
    |f({\bf q} + {\bf G}_2)|^2 V^{(0)}({\bf q} + {\bf G}_2 )- V_{ii}
    \label{fin-Cou}
    \end{eqnarray}
where ${\bf G}_2 = 2\pi (n,m)/a$ is the reciprocal wave vector for
square lattice and
        \begin{eqnarray}
        V^{(0)}({\bf q}) &=& \frac{V}{qa \tanh(q/q_0)} \;,
        \label{continuum} \\
        V &=& \frac{2\pi e^2}{a\sqrt{\epsilon_{ab}\epsilon_c}} \;,
        \label{V} \\
        q_0&\equiv& 2/\tilde d=(2/d)\sqrt{\epsilon_c/\epsilon_{ab}} \;.
        \label{q0}
        \end{eqnarray}
Here $\tilde d$ is an effective interlayer distance. The potential
(\ref{continuum}) interpolates between 3D $Vq_0/q^2a$
and 2D $V/qa$ limits at small $q\ll q_0$ and large $q\gg q_0$ momenta,
respectively. This crossover at $q_0$ reflects the fact that the planes
are independent at large momenta. For La$_2$CuO$_4$ compound where
$\epsilon_c/\epsilon_{ab} \sim 1/2$, \cite{Rea89} the value of crossover
momentum is estimated as $q_0\simeq 0.8/a$.

The function (\ref{fin-Cou}) is explicitly periodic in $q-$space.
In the particular model for the formfactor, Eq.\ (\ref{fq}), the
subtracted term $V_{ii}$ is evaluated as $ V_{ii} = 3 \kappa a V /32$ .
In general, $V_ {\bf q}$ is sign-reversal function in the Brillouin
zone,
as it should be in view of $\sum_{\bf Q} V({\bf Q}) = V({\bf R}=0)=0$.
In case of a continuum limit for the planes \cite{Fet74} (instead of the
tight-binding model used here), the Coulomb potential would be given
solely by Eq.(\ref{continuum}), the result formally obtained from
Eq.(\ref{fin-Cou})
by setting $V_{ii}=0$, $f({\bf q})=1$ and taking ${\bf G}_2=0$ term
alone. Finally, we note for the completeness
that $V({\bf q},q_z)$ for arbitrary $q_z$ is again given by
Eq.(\ref{fin-Cou}) but the potential $V^{(0)}({\bf q})$ for
a contimuum limit in Eq.(\ref{continuum}) must be
replaced by $V^{(0)}({\bf q},q_z) = V^{(0)}({\bf q})/(1+F^2_z)$,
where $F_z=\sin(q_z d/2)/\sinh(q\tilde d/2)$. \cite{Fet74}

At moderate $\kappa \sim 1$ {\em and} small momenta, $q<\kappa$,
(or, alternatively, at distances exceeding the size of Zhang-Rice
singlet) only ${\bf G}_2=0$ term in (\ref{fin-Cou}) contributes and
$V_ {\bf q} \simeq V^{(0)}({\bf q})$. The rapid decay of
$|f({\bf q})|^2$, (\ref{fq}), cuts off the values of
$V_ {\bf q}$ at larger momenta, $q\agt \kappa$.

The opposite limit, $\kappa a \gg 1$, corresponds to the point-charge
approximation and the sum in (\ref{fin-Cou}) formally diverges with
$\kappa$. This divergence is cancelled by the above on-site term
$V_{ii}\propto \kappa$. In Appendix, we provide another representation
for $V_{\bf q}$ in the point-charge limit; it works increasingly well
for $\kappa a \agt 3$.

\begin{figure}[tp]
\includegraphics[width=8cm]{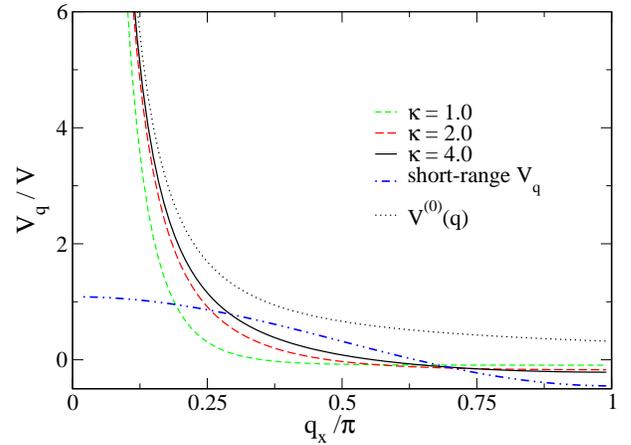}
\caption{
Coulomb potential $V_{\bf q}/V$ for different values of $\kappa$.
A crossover momentum $q_0=0.8/a$. For comparison, a dashed-dotted line
shows
a simple model which includes NN- and NNN-repulsion only. Dotted line
is calculated from Eq.(\ref{continuum}) neglecting the lattice structure
within the planes.
\vspace*{0.5cm}
\label{fig:6}}
\end{figure}

Fig.\ \ref{fig:6} shows ${\bf q}$-dependence of the Coulomb potential,
Eq.(\ref{fin-Cou}),
as function of parameter $\kappa$ at fixed $q_0=0.8/a$.
$V_{\bf q}/V$ evolves from the point-charge limit with visible
sign-reversal character to nearly positive curves as $\kappa$ decreases.
For comparison, we also show a frequently used simple short-range
repulsion model,
 \[  V_q = 2V_1(\cos q_x +\cos q_y) + 4V_2 \cos q_x \cos q_y, \]
with $V_1=V/2\pi, V_2=V_1/\sqrt{2}$
(dashed-dotted line), and the result that would be obtained when
lattice structure is discarded within the planes (dotted line). The
latter
is always positive as should be for the charges in a continuum.

Let us turn now to the charge susceptibility and consider how it is
influenced by long-range Coulomb interactions. First, we estimate
the energy scale $V$ in Eq.(\ref{V}). Using a representative value
$\sqrt{\epsilon_{ab}\epsilon_c}\sim 30$, \cite{Rea89}
one finds $V\simeq 0.8$~eV, which is about $2$ in units
of $t$. \cite{note0} Second, we should in principle complement the
Coulomb
potential with a short-range interactions between the holes, stemming
from local physics. One such a contribution is that of well-known
"missing $J$-link", which gives NN-attraction of the scale of
$J \langle {\bf s}_i{\bf s}_j-1/4 \rangle \sim -(0.1\div 0.2)t$. Yet
another local interaction is mediated by the bond-stretching vibration
of an oxygen shared by the two NN-holes. This contribution is repulsive
at low energy limit because of the coupling geometry (see for details
the next section), and is given by a half of the
polaron binding energy $E_b/2\sim t/4$ (estimated below from the phonon
shift induced by doping). Altogether, these two local NN-contributions
of different sign tend to cancel each other and their small net result
could be neglected. Hence, we focus on the Coulomb repulsion.

\begin{figure}[tp]
\includegraphics[width=8cm]{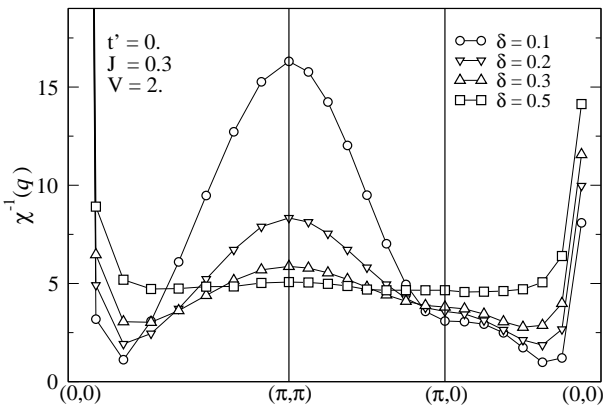}
\caption{
Behavior of inverse charge  susceptibility
at $J=0.3$, $t'=0$. Coulomb repulsion with $V=2$, $\kappa=2/a$ and
$q_0=0.8/a$ is included. Compare with figure 5.
\label{fig:7}}
\end{figure}

Fig.\ \ref{fig:7} shows a momentum dependence of the charge
susceptibility
at the presence of Coulomb interactions. The parameters used:
$q_0=0.8/a$ and $V=2$. Compared with a pure $t$-$J$
model result in Fig.4, one observes that $V_{\bf q}$ eliminates
phase separation effects. We find also that small momentum divergences
of the pairing fluctuations are suppressed, too. However, Coulomb
interaction
effects are not significant at larger momenta. All the local correlation
effects, which lead to a pronounced anisotropy of the charge
susceptibility
and its nontrivial doping dependence along the $\Gamma - X$ direction,
remain intact. As expected, the main effect of Coulomb interactions is
to move a potential charge instabilities to a finite momenta,
as seen in Fig.\ \ref{fig:7}. The divergence of $\chi^{-1}_{\bf q}$ at
$q\to0$ corresponds to $\chi({\bf q}=0)=0$, a well-known screening
phenomenon in the presence of long-range Coulomb repulsion.

\section{application: phonon softening in cuprates}

Density fluctuations determine doping induced phonon renormalization,
as discussed earlier in a framework of slave-boson
method, \cite{Kha97,Hor05} and also by an exact diagonalization
of small clusters. \cite{Ros04a}
In particular, a broad and anomalous lineshape of the bond-streching
phonons with momentum at $(\pi,0)$ direction has been found,
\cite{Kha97}
while no such anomalies were present for $(\pi,\pi)$. The effect is
strongly
doping dependent. \cite{Hor05} We consider now how the phonon
renormalization
effects are modified when the pairing (pseudogap formation) is included.
Specifically, we address here a phonon softening problem and discuss the
results in the context of experimental reports. \cite{Pin05,Fuk05,Rez05}

A doped hole couples to the bond-stretching vibrations of its four
oxygen neighbours \cite{Kha97}:
\begin{equation}
{\cal H}
=g \sum_i n_i (u^i_x-u^i_{-x}+u^i_y-u^i_{-y}).
\end{equation}
In a momentum space this reads as
\begin{equation}
{\cal H}
=i \sqrt{E_b\omega_0} \sum_{{\alpha,\bf q}}
\sin(q_{\alpha}a/2) (a_{\alpha,-{\bf q}} + a^{\dagger}_{\alpha,{\bf q}})
n_{{\bf q}}.
\label{He-ph}
\end{equation}
Here, $\alpha=x$ and $y$  denotes polarization of the oxygen
displacement,
and $\omega_0=\sqrt{K/m}$ is the phonon frequency determined by spring
constant $K$ and the oxygen mass $m$. Electron-phonon coupling strength
is conveniently quantified by a binding energy
\begin{equation}
E_b=2g^2/K ,
\end{equation}
that would be gained in case of a static hole. By fitting
a slave-boson theory to the experimental data on phonon softening
and linewidth in cuprates, an estimation $E_b\sim t/2$
was obtained in Ref.\ \onlinecite{Kha97}.

We assume that the energies of density fluctuations
are higher than the phonon energy --- it should be valid
not too close to charge instability. Within this assumption,
we can use our static $\chi_{{\bf q}}$ to estimate phonon softening,
$\delta\omega_{{\bf q}}=\omega_0-\omega_{{\bf q}}(\delta)$,
which is obtained as follows:
\begin{equation}
\frac{\delta \omega_{{\bf q}}}{\omega_0}=
1-\sqrt{1-2E_b\chi_{{\bf q}}(1-\gamma_{{\bf q}})} \\
\simeq E_b\chi_{{\bf q}}(1-\gamma_{{\bf q}}) ,
\label{phonon}
\end{equation}
where $\gamma_{{\bf q}}=(\cos q_x+\cos q_y)/2$ and
$\chi_{\bf q}$ includes the Coulomb repulsion.
Note also, that without correlations the product
$E_b\chi^{(0)}_{{\bf q}}\simeq E_b/4t$. One can therefore introduce
a dimensionless quantity, $\lambda^{(0)}=E_b/4t$, which could be
regarded as a "bare" coupling constant in the problem. According
to Ref.\ \onlinecite{Kha97}, $\lambda^{(0)}\simeq 1/8$,
justifying a perturbative treatment. Eq.\ (\ref{phonon}) reads now as
\begin{eqnarray}
\delta \omega_{{\bf q}}/\omega_0 &=& \lambda_{{\bf q}} =
\lambda^{(0)}\;4t\chi_{\bf q}(1-\gamma_{{\bf q}}).
\label{lambda}
\end{eqnarray}

Factor $\gamma_{{\bf q}}$ in Eq.(\ref{lambda}),
which stems from the coupling geometry,
would suggest a strongest softening for the full-breathing mode,
that is at ${\bf q}=(\pi,\pi)$. However, strong correlations change a
momentum dependence of $\chi_{{\bf q}}$ dramatically,
by suppressing it at $(\pi,\pi)$
and enhancing around $(\pi,0)$ points. One may say that correlations
lead to the redistribution of the effective electron-phonon coupling
in a momentum space. As a result, softening becomes
strongest at $(\pi,0)$, consistent with experiment. This explanation
of Ref.\ \onlinecite{Kha97} is further supported by the present
calculation, including the pseudogap and $t'$ effects.

\begin{figure}[tp]
\includegraphics[width=8cm]{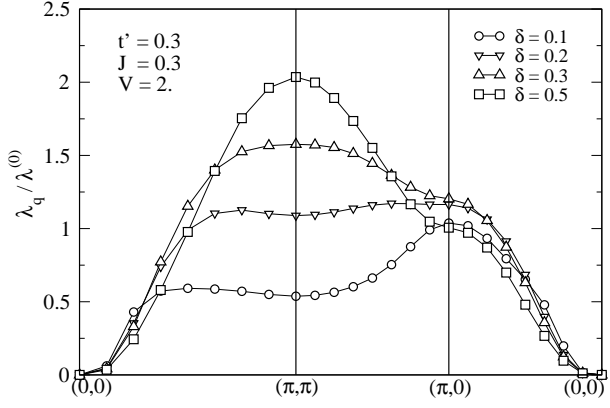}
\caption{
Momentum dependence of the effective coupling constant (in units of the
bare coupling $\lambda^{(0)}$) which determines renormalization of the
bond-stretching phonons. Parameters used: $t'=0.3$, $J=0.3$,
$V=2$ and $\kappa=2/a$.
\label{fig:8}}
\end{figure}

Fig.\ \ref{fig:8} shows a general form of the renormalization factor
$\lambda_{{\bf q}}/\lambda^{(0)}$ along particular directions
in the Brillouin zone. This figure is in obvious correspondence
with the above findings, and particularly, with Fig.\ \ref{fig:7}.

\begin{figure}[tp]
\vspace*{0.5cm}\includegraphics[width=8cm]{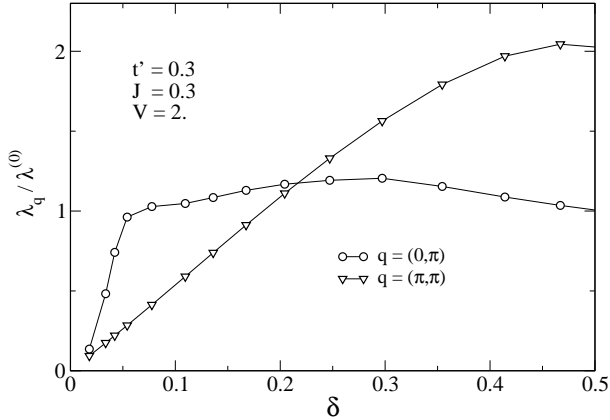}
\caption{
Renormalization of phonons (in units of the bare coupling
constant $\lambda^{(0)}$) at the symmetry points as a function
of doping.
The parameters used are
as in Fig.\ \ref{fig:8}.
\label{fig:9}}
\end{figure}

Fig.\ \ref{fig:9} presents more detailed doping
dependence of $\lambda_{{\bf q}}/\lambda^{(0)}$ at ${\bf q}=(\pi,\pi)$
and ${\bf q}=(\pi,0)$.
When multiplied by a bare constant $\lambda^{(0)}$, these curves
correspond to the phonon softening $\delta\omega_{{\bf q}}/\omega_0$.
As the latter is about 15 -- 20\% for $(\pi,0)$ phonon in optimally
doped
cuprates, \cite{Pin05,Fuk05} a bare constant $\lambda^{(0)}$ of the
order
of 0.15 -- 0.20 is required to fit the observed data \cite{notelambda}.
A striking similarity with the observed doping dependence
\cite{Pin05,Fuk05}
is worth to be pointed out here: both in experiment and in our theory
phonon softening along $(\pi,0)$ direction is almost independent
on doping in a wide region above $\delta \sim 0.12$.
While such a trend was already found earlier, \cite{Hor05}
$J$-pseudogap effects dramatically enhance the charge susceptibility
along $(\pi,0)$ at small $\delta$, hence it obtains nearly flat
doping dependence rather unexpected in view of hole-dilution physics.

\begin{figure}[tp]
\vspace*{0.4cm}\includegraphics[width=8cm]{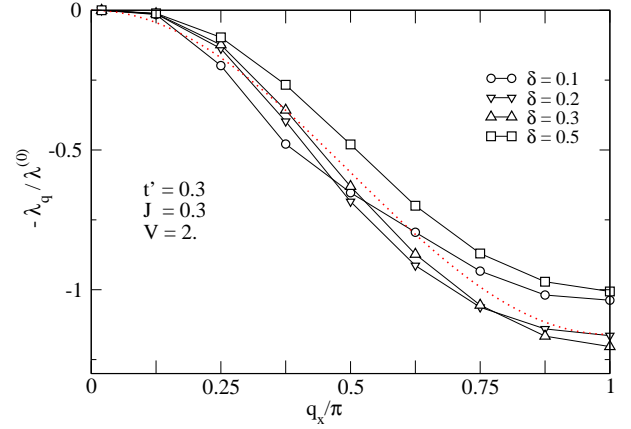}
\caption{
Renormalization of phonons along the $(0,\pi)$ direction
(in units of $\lambda^{(0)}$). The parameters used: $t'=0.3$, $J=0.3$,
$V=2$, $\kappa=2/a$ and $q_0=0.8/a$. A simple cosine curve is shown
for comparison by dotted line.
\label{fig:10}}
\end{figure}

For further comparison of our theory with the available experimental
data,
we plot a momentum dependence of $\lambda_{{\bf q}}/\lambda^{(0)}$
along $\Gamma - X$ direction for several values of doping
in Fig.\ \ref{fig:10}. One finds that visible deviations from a simple
cosine curve increase at smaller dopings, in general agreement with
experiment. \cite{Fuk05,Rez05}

While present calculations do capture the most anomalous experimental
findings -- stronger and nonlinear doping effects along ($\pi,0$)
direction -- a quantitative comparison is much less satisfactory.
In particular, rather sharp kink-like change in doping dependence
is observed in experiment at about $\delta \sim 0.12$,
\cite{Fuk05,Pin05} while it is found in our theory at lower doping.
One obvious reason for this discrepancy is that, strickly speaking,
we cannot quantitatively address the phonon softening
problem by using our {\it static} charge susceptibility. This is because
the charge fluctuations for momenta along $(\pi,0)$ direction
extend to low energies (comparable to phonon ones), as shown both in a
slave-boson
theory \cite{Kha96} and in the numerical work. \cite{Toh95,Ede95}
A dynamical susceptibility is therefore required which is however
beyond the scope of present work. Yet another reason is that, focusing
mainly on the pairing effects, we did not include an effective hopping
contribution stemmimg from ``Fock'' decoupling of the $J$ interaction.
This contribution, which is a fraction of $J/t$, \cite{Wan91,Kha96} will
stabilize a Gutzwiller band narrowing effect at doping levels
$\sim (0.2-0.3) J/t$ below which a linear doping dependence
$\chi_{\vec q} \propto \delta$ due to a hole-dilution effect sets in.
This is expected to shift a kink feature to higher dopings as in
experiment. Further, a quantitative description would
also require the inclusion
of spin-polaron effects beyond the leading $1/N$
approximation \cite{Kha96,Hor05}, and a coupling of the charge
fluctuations to a collective spin mode. \cite{Hor06}

In general, it seems that a kink feature in the doping dependence
of $(\pi,0)$ phonon softening \cite{Pin05,Fuk05}, which apparently
parallels with the so-called ``Yamada plot'' for the spin
incommensurability \cite{Yam98}, provides an interesting test case
for theory. Its initial linear behavior $\propto \delta$ is expected
and can be explained in terms of sum rule arguments. \cite{Ros04}
The saturation above certain doping level is well captured by
slave-boson
theories, but it is not fully clear at present why such an abrupt
regime change happens at around the ``magic'' doping $\sim 1/8$
concomitant with the saturation of spin incommensurability.

Finally, it is interesting to notice that the renormalization
of density fluctuations and the so-called charge vertex
$\gamma(k,q)$ of Refs.\onlinecite{Kul94,Zey96} (denoted by $\Gamma(p,q)$
in Refs.\onlinecite{Hua03,Koc04,Cit05}) have a common origin. Reflecting
this, the charge susceptibility can readily be expessed
via $\gamma(k,q)$ (see, {\it e.g.} Eq.(16) of Ref.\onlinecite{Zey96}).
In fact, in case of zero-frequency and small momenta one finds
$\gamma(k_f,{\bf q})\simeq \delta \cdot (\chi_{{\bf q}}/\chi^{(0)}_{{\bf
q}})$.
This relation makes it clear the origin of strong momentum structure
in electron-phonon vertex function found in
Refs.\onlinecite{Kul94,Zey96,Bec96,Hua03,Koc04,Cit05} --- this simply
reflects
highly momentum-selective action of the Gutzwiller constraint on density
fluctuations as we emphasized in previous sections of this paper.
Indeed, as $\chi_{{\bf q}}/\chi^{(0)}_{{\bf q}}\propto 1/\delta$ at
small
momenta, one realizes that the effective electron-phonon interaction is
essentially the bare one. At large momenta, however, it is strongly
suppressed as the hole density $\delta$ is reduced. \cite{note1}
This results in a "predominantly forward scattering" of electrons on
phonons in cases when this coupling is located in a density channel
(which is the case for the scattering on bond-stretching phonons).
In the context of cuprates, one should however realize that this
small-momentum peak structure in $\chi_{{\bf q}}$, hence also in
$\gamma(k,q)$
is in fact suppressed by long-range Coulomb repulsion. Moreover,
a bare matrix element for bond-stretching phonons is itself vanishing
as $q^2$ at small ${\bf q}$ [observe the matrix elements
$\sin(q_{\alpha}a/2)$ in
Eq.(\ref{He-ph}) and resulting form-factor $(1-\gamma_{{\bf q}})$
in Eq.(\ref{phonon})]. These two factors eliminate the
"forward-scattering" feature in cuprates. Altogether,
it seems that effective coupling of the Fermi-surface electrons
to bond-stretching phonons is somewhat reduced from $\lambda^{(0)}$
for both small and large values of momentum-transfer ${\bf q}$.
Therefore, a significance of the bond-streching phonons for the
electronic properties of doped cuprates should not be
overemphasized. \cite{note1}

\section{conclusions}

We calculated a momentum dependence of the static charge susceptibility
in the $t-t'-J$ model at various doping levels. We employed the
formalism, which effectively resums the RPA series in Gutzwiller,
$J$-term and Coulomb repulsion $V$-channel simultaneously.
We observe that
$\chi({\bf q})$ is a featureless
function in the limit of weakly correlated overdoped regime.
With decreasing of the doping level, strong correlations lead to a
nontrivial, highly momentum-dependent renormalization
effects that cannot be described
in terms of simple hole-density dilution effects. We demonstrate that
$\chi({\bf q})$ is strongly
suppressed near $(\pi,\pi)$. However, $\chi({\bf q})$ remains large
around $(\pi,0)$ and $(0,\pi)$ regions even at dopings as small as
0.1--0.2. A strong anisotropy of charge dynamics at finite wave-vectors
has
important experimental consequences, leading, {\it e.g.} to a pronounced
anisotropy in a renormalization of the bond-stretching phonon
modes as observed in cuprates.
We find that the exchange $J$ and second hopping $t'$ influence
the charge susceptibility in the opposite way -- the former enhances
it while $t'$ and also Coulomb interactions weaken
the small-momenta anomalies in $\chi({\bf q})$.
Implications of these findings on
the spin and electronic properties deserve further analysis.
The present study specifies the most "dangerous" regions
in a momentum space where one may expect charge-related anomalies.
A complex interplay between the charge, spin and fermionic excitations
should be considered to fuller extent in order to locate more precisely
a momentum position of low-energy charge modulations.

\begin{acknowledgments}
We would like to thank I.~Gornyi, O.~Gunnarsson, P.~Horsch,
O.P.~Sushkov and M.~Vojta for useful discussions and comments.

\end{acknowledgments}

\appendix* \section{Coulomb interaction on the lattice: point-charge
limit}

Consider the formula for the Coulomb interaction
(\ref{Vij}) for $f({\bf r}) = \delta({\bf r})$.
Introducing an auxiliary integration, the Fourier transform
$ V_{\bf q}$ at $q_z=0$ can be written as
     \begin{equation}
     V_{\bf q} =
     \frac{V}{\pi^{3/2} }
     \int_0^\infty d\tau
     \sum_{{\bf r}_i\neq 0}
     e^{-\tau^2 (r_i^2 + \tilde z_i^2)
     +i{\bf q}{\bf r} }
     \label{auxV}
     \end{equation}
where $r_i^2=x_i^2 + y_i^2$ denote square lattice sites with
$x_i, y_i = (0, \pm 1, \pm 2, \ldots)$, while
$\tilde z_i=(0, \pm 1, \pm 2, \ldots) \tilde d$ ; we set $a=1$ here.
We use now the definition of the Jacobi $\vartheta$ function,
\[\vartheta_3(u, q) = \sum_{n=-\infty}^\infty  q^{-n^2} e^{2iun} ,\]
and represent Eq.\ (\ref{auxV}) in the form
     \begin{eqnarray}
     V_{\bf q} &=&
     \frac{V}{\pi^{3/2} }
     \int_0^\infty d\tau
     \left[
     \varphi\left(\frac{ {\bf q}}2, e^{-\tau^2 } \right) -1 \right],
     \label{Coulomb-points}
     \\
     \varphi({\bf q}, s) &=& \vartheta_3\left( q_x , s  \right)
     \vartheta_3\left( q_y , s \right)
     \vartheta_3\left( 0 , s^{\tilde d^2} \right).
     \end{eqnarray}
This formula can be considered as the limiting case of
Eq.~(\ref{fin-Cou})
with $\kappa \to \infty$. Numerically, the values of $V_{\bf q}$ in the
whole Brillouin zone given by Eqs.~(\ref{fin-Cou}) and
(\ref{Coulomb-points}) become very close at $\kappa \agt 3$.



\begin{thebibliography}{}

\bibitem{Han04} T.~Hanaguri, C.~Lupien, Y.~Kohsaka, D.-H.~Lee,
M.~Azuma, M.~Takano, H.~Takagi, and J.C.~Davis,
Nature {\bf 430}, 1001 (2004).

\bibitem{Pin05} L.~Pintschovius,
Phys.~Stat.~Sol.~(b), {\bf 242}, 30 (2005).

\bibitem{Fuk05} T.~Fukuda, J.~Mizuki, K.~Ikeuchi, K.~Yamada,
A.Q.R.~Baron,  and S.~Tsutsui,
Phys.~Rev.~B {\bf 71}, 060501 (2005).

\bibitem{Rez05} D.~Reznik, L.~Pintschovius, M.~Ito, S.~Iikubo, M.~Sato,
H.~Goka, M.~Fujita, K.~Yamada, G.D.~Gu, and J.M.~Tranquada,
Nature {\bf 440}, 1170 (2006). 

\bibitem{Wan91} Z.~Wang, Y.~Bang, and G.~Kotliar,
Phys.~Rev.~Lett. {\bf 67}, 2733 (1991).

\bibitem{Geh95} L.~Gehlhoff and R.~Zeyher,
Phys.~Rev.~B {\bf 52}, 4635 (1995).

\bibitem{Kha96} G.~Khaliullin and P.~Horsch,
Phys.~Rev.~B {\bf 54}, R9600 (1996).

\bibitem{Zim97} W.~Zimmermann, R.~Fr\'esard, and P.~W\"olfle,
Phys.~Rev.~B {\bf 56}, 10097 (1997).

\bibitem{Toh95} T.~Tohyama, P.~Horsch, and S.~Maekawa,
Phys.~Rev.~Lett. {\bf 74}, 980 (1995).

\bibitem{Ede95} R.~Eder, Y.~Ohta, and S.~Maekawa,
Phys.~Rev.~Lett. {\bf 74}, 5124 (1995).

\bibitem{Kim91} J.H.~Kim, K.~Levin, R.~Wentzcovitch, and A.~Auerbach,
Phys.~Rev.~B {\bf 44}, 5148 (1991).

\bibitem{Kul94} M.L.~Kuli\'c and R.~Zeyher,
Phys.~Rev.~B {\bf 49}, 4395 (1994); R.~Zeyher and M.L.~Kuli\'c,
{\it ibid.} {\bf 53}, 2850 (1996).

\bibitem{Zey96} R.~Zeyher and M.L.~Kuli\'c,
Phys.~Rev.~B {\bf 54}, 8985 (1996).

\bibitem{Bec96} F.~Becca, M.~Tarquini, M.~Grilli, and C.~Di Castro,
Phys.~Rev.~B {\bf 54}, 12443 (1996).

\bibitem{Koc04} E.~Koch and R.~Zeyher, Phys.~Rev.~B {\bf 70}, 094510
(2004).

\bibitem{Hua03} Z.B.~Huang, W.~Hanke, E.~Arrigoni, and D.J.~Scalapino,
Phys.~Rev.~B {\bf 68}, 220507(R) (2003).

\bibitem{Ros04} O.~R\"osch and O.~Gunnarsson,
Phys.~Rev.~Lett. {\bf 93}, 237001 (2004).

\bibitem{Cit05} R.~Citro, S.~Cojocaru, and M.~Marinaro,
Phys.~Rev.~B {\bf 72}, 115108 (2005).

\bibitem{Kha97} G.~Khaliullin and P.~Horsch,
Physica C {\bf 282-287}, 1751 (1997).

\bibitem{Hor05} P.~Horsch and G.~Khaliullin,
Physica B {\bf 359-361}, 620 (2005).

\bibitem{Fur92} N.~Furukawa and M.~Imada,
J.~Phys.~Soc.~Jpn. {\bf 61}, 3331 (1992).

\bibitem{Koh97} M.~Kohno,
Phys.~Rev.~B {\bf 55}, 1435 (1997).

\bibitem{Tan99}  A.~Tandon, Z.~Wang, and G.~Kotliar,
Phys.~Rev.~Lett. {\bf 83}, 2046 (1999).

\bibitem{McK01} R.H.~McKenzie, J.~Merino, J.B.~Marston, and
O.P.~Sushkov,
Phys.~Rev.~B {\bf 64}, 085109 (2001).

\bibitem{Hel97} M.S.~Hellberg and E.~Manousakis,
Phys.~Rev.~Lett. {\bf 78}, 4609 (1997);
see also E.~Dagotto, Rev.~Mod.~Phys. {\bf 66}, 763 (1994) and references
therein.

\bibitem{Dee94} M.~Deeg and H.~Fehske,
Phys.~Rev.~B {\bf 50}, 17874 (1994).

\bibitem{Bec00} A question as whether the compressibility is truly
divergent
is somewhat controversial: {\em e.g.}, F.~Becca,
M.~Capone and S.~Sorella [Phys.~Rev.~B {\bf 62}, 12700 (2000)] argue
that
the ground state is uniform.

\bibitem{Win98} S.~Winterfeldt and D.~Ihle,
Phys.~Rev.~B {\bf 58}, 9402 (1998).

\bibitem{Lar01} A.~Larkin, A.~Varlamov,
{\em Theory of Fluctuations in Superconductors} (International Series
of Monographs on Physics 127. Oxford U. Press, New York, 2005);
see also cond-mat/0109177.

\bibitem{Voj99} M.~Vojta and S.~Sachdev,
Phys.~Rev.~Lett. {\bf 83}, 3916 (1999).

\bibitem{Voj00} M.~Vojta, Y.~Zhang, and S.~Sachdev,
Phys.~Rev.~B {\bf 62}, 6721 (2000);
M.~Vojta, Phys.~Rev.~B {\bf 66}, 104505 (2002).

\bibitem{Pav01} E.~Pavarini, I.~Dasgupta, T.~Saha-Dasgupta,
O.~Jepsen, and O.K.~Andersen,
Phys.~Rev.~Lett. {\bf 87}, 047003 (2001).

\bibitem{Toh04} T.~Tohyama,
Phys.~Rev.~B {\bf 70}, 174517 (2004).

\bibitem{Kot04} V.N.~Kotov and O.P.~Sushkov,
Phys.~Rev.~B {\bf 70}, 195105 (2004).

\bibitem{Whi99} S.~White and D.J.~Scalapino,
Phys.~Rev.~B {\bf 60}, R753 (1999).

\bibitem{Rea89} D.~Reagor, E.~Ahrens, S-W.~Cheong, A.~Migliori, and
Z.~Fisk,
Phys.~Rev.~Lett. {\bf 62}, 2048 (1989).

\bibitem{Fet74} A.L.~Fetter,
Ann.~Phys.~(N.Y.) {\bf 88}, 1 (1974).

\bibitem{note0} We note by passing that the present model estimates
the in-plane plasma frequency as $\omega_p\simeq
\sqrt{V_{\infty}t\delta}$,
where $V_{\infty}=4\pi e^2/d\epsilon_{\infty}$. With a high-frequency
screening constant $\epsilon_{\infty}\simeq 5$, \cite{Rea89} and
$t=0.43$~eV, \cite{Pav01} this gives $\omega_p\simeq 0.7$~eV at
$\delta=0.2$, close to the observed value $\sim$ 0.8 eV. \cite{Uch91}

\bibitem{Uch91} S.~Uchida, T.~Ido, H.~Takagi, T.~Arima, Y.~Tokura,
and S.~Tajima,
Phys.~Rev.~B {\bf 43}, 7942 (1991).

\bibitem{Ros04a} O.~R\"osch and O.~Gunnarsson,
Phys.~Rev.~Lett. {\bf 92}, 146403 (2004).

\bibitem{notelambda} A smallness of this constant might seem surprising
in view of strong anticipated change of the Zhang-Rice singlet energy by
oxygen displacements. However, a doped hole
wave-function is rather extended and made predominantly of the oxygen
orbitals. \cite{Pav01} Hence, the energy-modulation by displacement of
the oxygen sites themselves could be not very large. In fact, LDA
calculations show that this coupling is indeed rather weak, see
S.Y.~Savrasov and O.K.~Andersen,
Phys.~Rev.~Lett. {\bf 77}, 4430 (1996);
O.K.~Andersen, S.Y.~Savrasov, O.~Jepsen, and A.I.~Liechtenstein,
J.~Low~Temp.~Phys. {\bf 105}, 285 (1996).

\bibitem{Hor06} P.~Horsch and G.~Khaliullin, unpublished.

\bibitem{Yam98} K.~Yamada, C.H.~Lee, K.~Kurahashi, J.~Wada, S.~Wakimoto,
S.~Ueki, H.~Kimura, Y.~Endoh, S.~Hosoya, G.~Shirane, R.J.~Birgeneau,
M.~Greven, M.A.~Kastner, and Y.J.~Kim,
Phys.~Rev.~B {\bf 57}, 6165 (1998).

\bibitem{note1} These findings are related to the scattering of
electrons in a metallic phase when large Fermi-surface is formed.
A single-hole problem is radically different: in that case,
correlations may considerably enhance the electron-phonon coupling constant,
see, {\it e.g.}, A.~Ram\v{s}ak, P.~Horsch, and P.~Fulde,
Phys.~Rev.~B {\bf 46}, 14305 (1992); A.S.~Mishchenko and N.~Nagaosa,
Phys.~Rev.~Lett. {\bf 93}, 036402 (2004). Moreover, the bare coupling constant
for a single-hole in insulator is itself different (larger and non-local)
since no metallic screening of the doped-charge is present in that case.

\end{thebibliography}
\end{document}